\documentclass[pdftex,twocolumn,epjc3]{svjour3}  
\smartqed  
\RequirePackage{graphicx}
\RequirePackage{flushend}
\RequirePackage[numbers,sort&compress]{natbib}
\RequirePackage[colorlinks,citecolor=blue,urlcolor=blue,linkcolor=blue]{hyperref}

\usepackage{graphicx}
\usepackage{dcolumn}
\usepackage{bm}
\usepackage{doi}
\def\mean#1{\left< #1 \right>}
\usepackage{multirow}
\usepackage{booktabs}
\usepackage{url}
\usepackage{hyperref}
\providecommand{\e}[1]{\ensuremath{\times 10^{#1}}}
  
\usepackage{savesym}
\savesymbol{tablenum}
\usepackage{siunitx}
\newcommand\solarmass{M\textsubscript{\(\odot\)}}
\journalname{Eur. Phys. J. C}
\usepackage[switch]{lineno}

\begin{document}

\title{Sensitivity of multi-PMT Optical Modules in Antarctic Ice to Supernova Neutrinos of MeV energy}

\author{C.~J.~Lozano Mariscal\thanksref{e1,addr1}\href{https://orcid.org/0000-0002-7159-493X}{\includegraphics[scale=0.5]{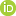}} \and 
L.~Classen\thanksref{addr1}\href{https://orcid.org/0000-0001-9495-5892}{\includegraphics[scale=0.5]{pictures/ORCIDiD_icon16x16.png}} \and
M.~A.~Unland~Elorrieta\thanksref{addr1}\href{https://orcid.org/0000-0002-6124-3255}{\includegraphics[scale=0.5]{pictures/ORCIDiD_icon16x16.png}} \and
A.~Kappes\thanksref{addr1}\href{https://orcid.org/0000-0003-1315-3711}{\includegraphics[scale=0.5]{pictures/ORCIDiD_icon16x16.png}}}

\institute{Institut f\"ur Kernphysik, Westf\"alische Wilhelms-Universit\"at M\"unster, Wilhelm-Klemm-Straße 9, 48149 M\"unster, Germany \label{addr1}
 }%

\thankstext{e1}{email: \href{mailto:c.lozano@wwu.de}{c.lozano@wwu.de} (corresponding author)}

\date{Received: 29 June 2021 / Accepted: 5 November 2021\\\textcopyright The Author(s) 2021}

\maketitle

\begin{abstract}
New optical sensors with a segmented photosensitive area are being developed for the next generation of neutrino telescopes at the South Pole. In addition to increasing sensitivity to high-energy astrophysical neutrinos, we show that this will also lead to a significant improvement in sensitivity to MeV neutrinos, such as those produced in core-collapse supernovae (CCSN). These low-energy neutrinos can provide a detailed picture of the events after stellar core collapse, testing our understanding of these violent explosions. We present studies on the event-based detection of MeV neutrinos with a segmented sensor and, for the first time, the potential of a corresponding detector in the deep ice at the South Pole for the detection of extra-galactic CCSN. We find that exploiting temporal coincidences between signals in different photocathode segments, a $27\ \mathrm{M}_{\odot}$ progenitor mass CCSN can be detected up to a distance of \SI{269}{kpc} with a false detection rate of \SI{0.01}{year^{-1}} with a detector consisting of \SI{10000}{} sensors. Increasing the number of sensors to \SI{20000}{} and reducing the optical background by a factor of ${\sim}140$ expands the range such that a CCSN detection rate of ${\sim}0.08$ per year is achieved, while keeping the false detection rate at \SI{0.01}{year^{-1}}.

\begin{description}
\item[Keywords:]
neutrino astronomy, neutrino telescopes, supernova neutrinos, segmented optical sensors
\end{description}
\end{abstract}

\section{\label{sec:introduction}Introduction}
A core-collapse supernova (CCSN) explosion is the final stage in the evolution of stars featuring masses greater than $\sim8$ solar masses (\solarmass)~\cite{Heger:2002cn}. Neutrinos play a crucial role during these explosions, carrying away most of the energy that is radiated during the collapse in a short burst of $\sim10\,\mathrm{s}$~\cite{Janka:2006fh,Mirizzi2016}. In 1987, the first and so far only neutrinos from a CCSN were detected. This supernova, named SN 1987A, exploded in the Large Magellanic Cloud, a satellite galaxy of the Milky Way at a distance of \SI{51.4}{kpc} from Earth, emitting a burst of neutrinos. During this episode, 25 neutrino events with estimated energies of \SI{\sim15}{MeV} were detected in temporal coincidence by three different neutrino observatories: twelve by the Kamiokande neutrino detector~\cite{1987original}, eight by the IMB detector~\cite{Bionta:1987qt} and five by the Baksan scintillation telescope~\cite{Alekseev:1988gp}. Even though this detection confirmed the general picture of CCSN explosions, more than 30 years later a detailed picture of the physics of the core collapse is still missing. 

Neutrino telescopes that use abundances of natural water or ice as detection medium such as Antares~\cite{ANTARES:2011hfw} in the Mediterranean Sea, Baikal~\cite{BAIKAL:1997iok} in Lake Baikal or IceCube~\cite{Aartsen:2016nxy} at the South Pole have studied neutrinos in recent decades. The deep ice at the South Pole, with its high transparency and low radioactive contamination, has proven to be an excellent site for the detection of high-energy atmospheric and astrophysical neutrinos (see e.g.\ \cite{Abbasi:2020jmh,IceCube:2018cha}). The IceCube Neutrino Observatory instruments one cubic kilometer of South Pole ice in depths between \SI{1450}{m} and \SI{2450}{m} with 5160 Digital Optical Modules (DOMs). DOMs are spherical glass pressure vessels equipped with one 10-inch photomultiplier tube (PMT) facing downwards and associated read-out electronics~\cite{Abbasi:2008aa}. The modules are mounted on strings with a horizontal spacing of \SI{125}{m} and a vertical distance of \SI{17}{m} between modules. Neutrinos are detected indirectly via secondary charged particles which are produced after a neutrino interacts in the ice or bedrock below the detector. During their passage, these particles induce Cherenkov light emission in the ice, which is detected by the photomultipliers.
 
The current string layout of IceCube is optimized to reconstruct the energy and direction of neutrinos with energies above \SI{\sim 100}{GeV}. The detection of light from MeV neutrinos by more than one module, however, the prerequisite for event-based reconstruction, is very unlikely, since the produced low-energy secondary particles travel only a few centimeters in the ice. Therefore, IceCube does not have the capability to detect MeV supernova neutrinos individually. Nevertheless, a nearby CCSN can be detected by the observation of an increased counting rate in all DOMs in a time window of $\sim10$ seconds, corresponding to the supernova neutrino burst~\cite{Abbasi:2011ss}. 

Based on the success of IceCube, new extensions to the original detector are being planned that will further improve its performance. The IceCube upgrade~\cite{Ishihara:2019aao} is scheduled for installation in the Australian summer of 2022/23 and consists of about 700 additional optical modules distributed across seven strings. This extension will significantly expand IceCube's sensitivity to neutrino oscillation parameters using atmospheric neutrinos. The construction of a large detector for high-energy astrophysical neutrinos, IceCube-Gen2~\cite{Aartsen:2020fgd}, is planned to begin in the second half of the decade and is designed to encompass a large volume with about \SI{10000}{} optical sensors on 120 strings. Beyond a mere increase in the number of sensors, novel module types with segmented photosensitive areas are being developed for IceCube-Gen2. It is expected that such modules will provide increased sensitivity not only for high-energy neutrinos, but also for MeV neutrinos. The segmentation will enable a new approach to study low-energy events using temporal coincidences between photons within the same module (local coincidences), improving the sensitivity to CCSN detection.

While closely spaced (unsegmented) sensors can also be used to detect CCSN neutrinos~\cite{Boser:2013oaa}, this scenario is not applicable to very sparsely instrumented neutrino telescopes like IceCube-Gen2, which are optimized for high-energy astrophysical neutrinos. The KM3NeT Collaboration~\cite{Adrian-Martinez:2016fdl} is developing a local-coincidence method for the segmented sensors of their sparse detectors in the Mediterranean Sea~\cite{Aiello:2021cot}. In this work, we investigate CCSN detection via local coincidences for a detector in the South Pole ice equipped with \SI{10000}{} multi-PMT digital Optical Modules (mDOMs)~\cite{Classen:2019tlb} which has been developed for the IceCube Upgrade~\cite{Ma:2020wmj}. 

The paper is organized as follows: in Section \ref{sec:mdom} an introduction to the mDOM concept and its features is provided. In Section \ref{sec:simulation} we describe the simulation of supernova neutrinos used for the sensitivity studies. The final Section \ref{sec:identication} describes a set of trigger conditions for MeV neutrino identification and background suppression, which is subsequently used to estimate the sensitivity of a detector equipped with mDOMs for the detection of extra-galactic CCSNe.

\section{\label{sec:mdom}A multi-PMT optical module for the South Pole ice}

\begin{figure}[t]
    \centering
        \includegraphics[width=\columnwidth]{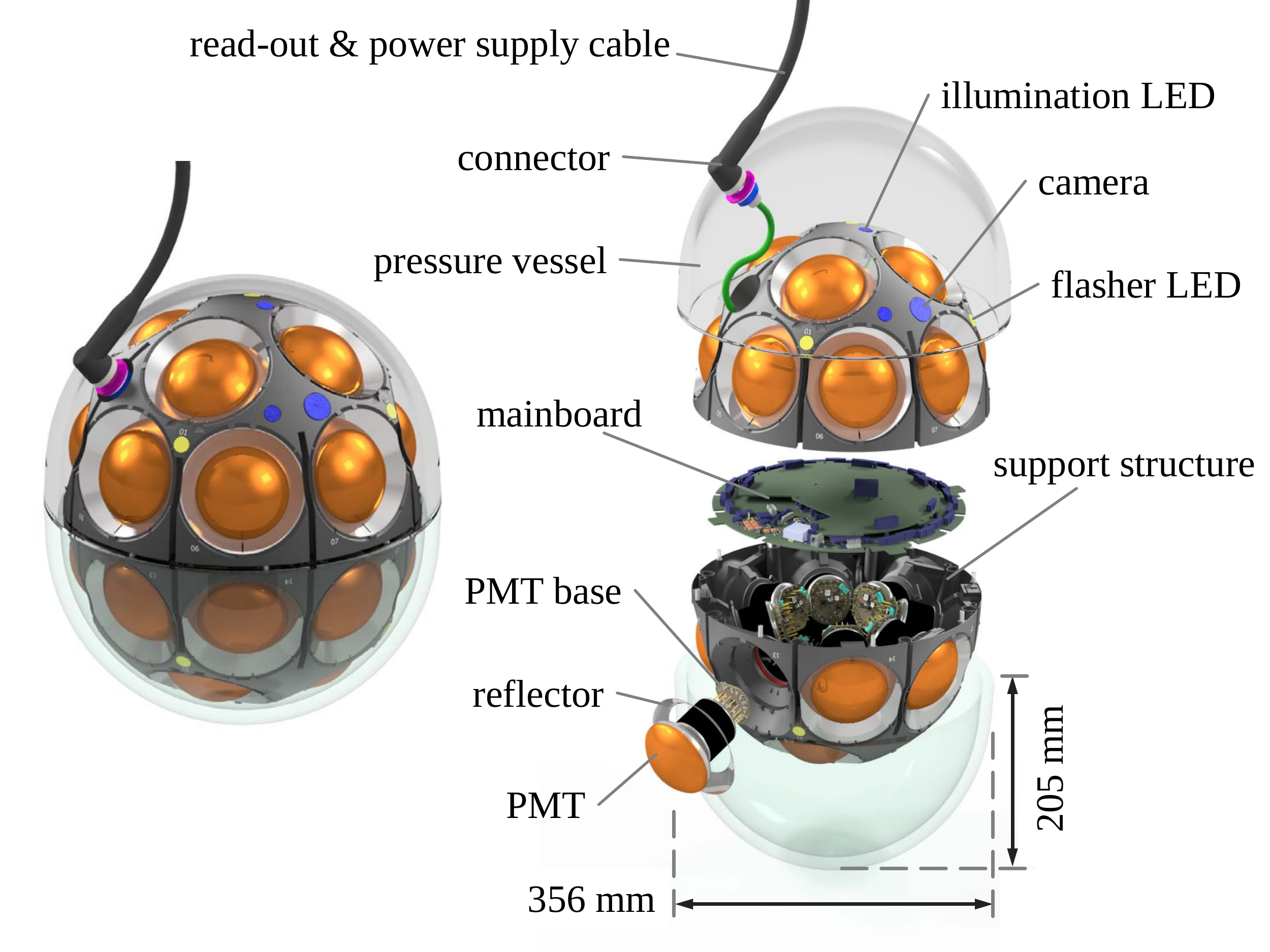}
    \caption{\textbf{Left:} Rendered picture of a mDOM. \textbf{Right:} Exploded view with labeling of the main components.}
    \label{fig:mDOM}
\end{figure}

The mDOM~\cite{Classen:2019tlb}, depicted in Fig.~\ref{fig:mDOM}, features 24 PMTs with 80$\,$mm photocathode diameter housed inside a pressure vessel. PMTs and support structure are coupled to the pressure vessel with optical gel. In contrast to the single large PMT in the current IceCube DOM, the PMTs in the mDOM cover the whole solid angle with high homogeneity. They are surrounded by reflector cones to further increase the PMT's effective area. This design results in several advantages:

\begin{itemize}
    \item Increased sensitive area per module by more than a factor two;
    \item A near uniform $4\pi$ coverage;
    \item PMT orientations provide intrinsic information about the photon direction, thereby improving event reconstruction;
    \item Simultaneous photon detection in several PMTs of the same module (local coincidence) which can be utilized to identify neutrino interactions over background.
\end{itemize}
In particular, the last point has the potential to distinguish photons from interacting MeV neutrinos from detector background. In this context, each module can be treated as an individual detector which performs an event-by-event detection. Thus, the sensitivity scales with the number of sensors and the results of our study are independent of the inter-module spacing.

\section{\label{sec:simulation}Simulation}

A GEANT4~\cite{Agostinelli:2002hh} Monte Carlo simulation of the mDOM \cite{lewthesis} has been adapted and is used to study the sensitivity of the module to MeV supernova neutrinos. The simulation takes into account the interactions of MeV neutrinos in the ice and the mDOM response, as well as the optical background produced by radioactive decays in the glass of the pressure vessel.

\subsection{\label{subsec:geant4}Detector simulation}
The mDOM simulation includes all relevant mechanical components of the module: pressure vessel, PMTs, reflector cones, support structure and optical gel. The PMT is modeled as a solid glass body with the photocathode on the inside. Each component has its own material properties. The support structure of the PMTs is simulated as a total absorbing massive object. When a photon reaches the surface of the photocathode it is removed and saved. Its detection probability is given by the quantum efficiency (QE) of the PMT. A further simulation of the PMT or front-end electronics is not included. The medium around the module was modeled after the properties of the South Pole ice~\cite{Aartsen:2013rt}. These vary with depth and have a layered structure reflecting cli\-ma\-to\-lo\-gi\-cal changes or volcanic activities in the past~\cite{Ackermann:2006pva}. In the simulation, absorption and scattering are parameterized as a function of the mDOM depth, however, on the scale of the simulated volume around the module, the ice properties are assumed to be constant. 

The simulations are performed using a single mDOM and afterwards are scaled to a detector with \SI{10000}{} modules. The varying ice properties in depths between \SI{1400}{} to \SI{2490}{m} \footnote{IceCube-Gen2 strings will probably be longer and extended into greater depths, increasing the number of modules in cleaner ice and thus the detector performance. However, in this work, the module distribution is limited to depths with measured ice properties by IceCube's LED calibration system \cite{Aartsen:2013rt}.} are taken into account by using the effective per-module volume (see section \ref{subsec:EffVol}). With a vertical module spacing of approximately \SI{13}{m} in the IceCube-Gen2 detector, shadowing effects can be neglected.

\subsection{\label{subsec:SNsimulation}SN neutrino burst}
The gravitational collapse of type II supernovae ends in a neutron star or, if the mass of the progenitor is larger than $\sim10\,$\solarmass, a stellar-mass black hole might be the outcome~\cite{O_Connor_2011}. During this collapse, electron capture by protons produces a temporally sharp electron neutrino burst lasting $\sim10\,\mathrm{ms}$, which is called \emph{neutronization peak}. After that, the main reaction that produces neutrinos and antineutrinos of all flavors is pair production. The neutrino energy spectra are parameterized as~\cite{Tamborra:2012ac,Keil:2002in}:
\begin{equation}
\label{eq:nuspectra}
    f_{\alpha(t)}(E_\nu,t)=E_\nu^{{\alpha}(t)} \cdot \exp \Big[-\big(\alpha(t)+1\big)E_\nu/\mean{E_\nu}(t)\Big],
\end{equation}
where $E_\nu$ and $\mean{E_\nu}$ are the neutrino energy and its mean, respectively, $t$ is the time after the collapse, and $\alpha(t)$ can be calculated from the moments of the neutrino energy spectrum,
\begin{equation}
\label{eq:alpha}
    \frac{\mean{E_{\nu}^2}(t)}{\mean{E_{\nu}}^2(t)}=\frac{2+\alpha(t)}{1+\alpha(t)}.
\end{equation}
In this work, we employ fluxes from two different CCSNe models~\cite{Sukhbold:2015wba}, which are based on the \emph{LS220} (Lattimer - Swesty) Equation of State (EoS)~\cite{Lattimer:1991nc}:
\begin{itemize}
    \item A CCSN with a progenitor mass of $27.0\,\mathrm{M_\odot}$, leading to a baryonic neutron star of $1.77\,\mathrm{M_\odot}$;
    \item A CCSN with a progenitor mass of $9.6\,\mathrm{M_\odot}$, leading to a baryonic neutron star of $1.36\,\mathrm{M_\odot}$.
\end{itemize}

\begin{figure}[t]
    \centering
    \includegraphics[width=\columnwidth]{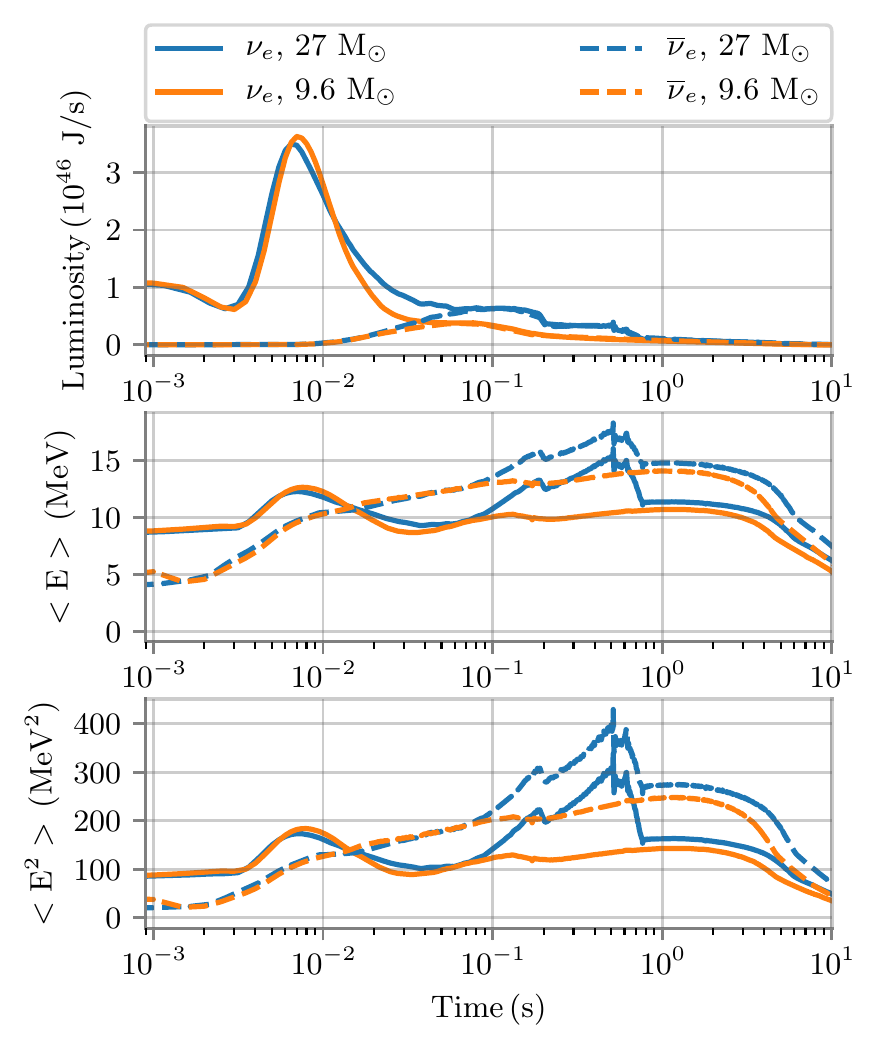}
    \caption{Time dependence of the luminosity, the mean energy and the mean squared energy of neutrinos and antineutrinos for the CCSNe models used in this work. Data taken from \cite{Sukhbold:2015wba}.}
    \label{fig:fluxes}
\end{figure}
The luminosity, mean energy, and mean squared energy as a function of time from the models are shown in Fig.~\ref{fig:fluxes}. 
\begin{figure}[t]
    \centering
    \includegraphics[width=\columnwidth]{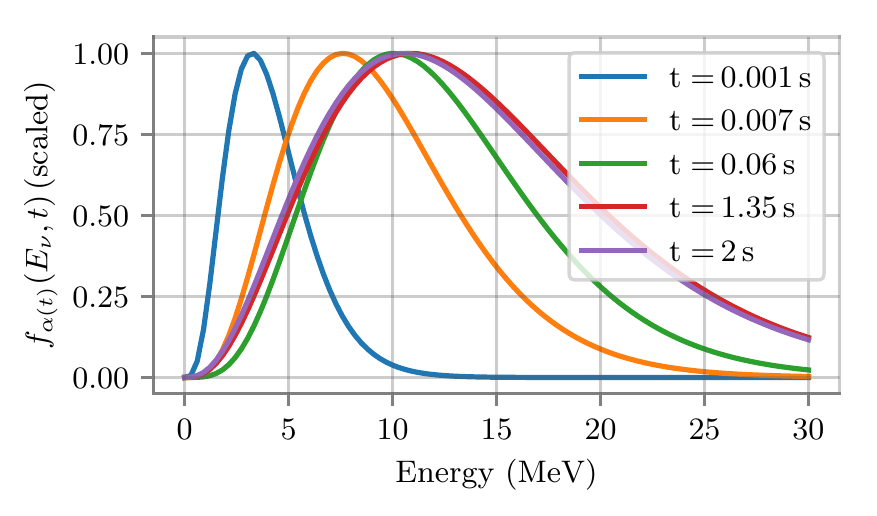}
    \caption{$\bar{\nu}_{e}$ energy spectrum of a CCSN with a $27.0\,\mathrm{M_\odot}$ progenitor mass at different times during the burst.}
    \label{fig:energydist}
\end{figure}

Upon reaching Earth, some neutrinos will interact within the instrumented volume. The most important reactions for a detector in the ice are elastic neutrino-electron scattering (ENES)
\begin{equation}
\label{eq:elasticscattering}
\nu_{e}+e^-\rightarrow\nu_{e}+e^- ,
\end{equation}
for electron neutrinos originating from the neutronization peak, and inverse beta decay (IBD)
\begin{equation}
\label{eq:inversebeta}
\bar{\nu}_{e}+p\rightarrow n+e^{+} ,
\end{equation}
which has the largest cross section of all MeV neutrino interactions in ice~\cite{Abbasi:2011ss,Vogel:1999zy}. The generated $e^\pm$ usually exceed the Cherenkov threshold and therefore produce visible light that propagates through the ice and can be detected by the optical modules. Note that we do not take into account interactions with oxygen nuclei or the scattering of other neutrino flavors, which reduces the event rate by about $5\%$ (see Table 1 in \cite{Abbasi:2011ss}).

The energy spectrum of the incoming neutrinos or antineutrinos can be calculated for each point in time using Eq. \ref{eq:nuspectra} and \ref{eq:alpha} with examples shown in Fig.~\ref{fig:energydist}. Higher energy neutrinos produce higher energy secondary particles in the ice, which in turn produce more photons, increasing the probability of being detected in one or more PMTs. Thus, although the mean energy of the CCSN neutrinos range mainly from about \SI{5}{} to \SI{15}{MeV} (see Fig.~\ref{fig:fluxes}), the larger effective cross section of the higher energy neutrinos and their higher Cherenkov photon yield lead to a mean energy of about \SI{25}{MeV} for the detected neutrinos.

Results shown in this work do not include any oscillation scenario, which is a conservative choice. In both normal and inverted mass ordering scenarios a larger number of detected events is expected than in the no-oscillation case, with the inverted scenario providing the largest signal because energetic $\bar{\nu}_x$
would oscillate into $\bar{\nu}_e$~\cite{Abbasi:2011ss}.

\subsection{\label{subsec:EffVol}Effective volume}
The effective detection volume of a mDOM for MeV neutrinos is calculated for different depths with the simulation described in Section~\ref{subsec:geant4}. Electrons are injected homogeneously into a sufficiently large spherical volume of South Pole ice surrounding the module. As they travel through the ice, they induce Cherenkov radiation that can be detected by the module's PMTs. The effective volume is given by
\begin{equation}
\label{eq:effvolume}
V_{\mathrm{eff}} = \frac{N_{\mathrm{det}}}{N_{\mathrm{gen}}}\cdot V_{\mathrm{gen}},
\end{equation}
where $N_{\mathrm{det}}$ and $N_{\mathrm{gen}}$ are the number of detected and generated electrons, respectively, and $V_{\mathrm{gen}}$ is the generation volume. Electrons producing at least one photon registered by a PMT are considered detected. A sufficiently large generation volume means that an increase in the generation volume does not increase the effective volume. To reduce computational effort, the mDOM is assumed to be spherically symmetric, requiring particles to be simulated only from one direction. This approximation is sufficient for the re-scaling of the effective volume used later in the analysis. The same light production is assumed for electrons and positrons.

\begin{figure}[t]
    \centering
    \includegraphics[width=\columnwidth]{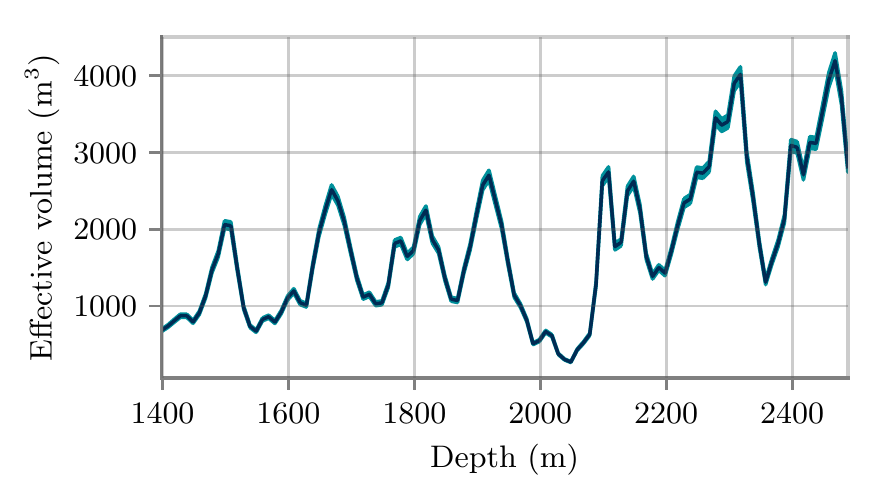}
    \caption{Effective detection volume of an mDOM for electrons with an energy of $\SI{25}{MeV}$ as a function of depth.}
    \label{fig:effvol}
\end{figure}

Simulations at different depths have been performed for an electron energy of \SI{25}{MeV}, where each depth has a corresponding absorption and scattering length taken from~\cite{Ackermann:2006pva}. The effective volume as a function of the module depth is shown in Fig.~\ref{fig:effvol}. 

\begin{table*}[t]
\centering
\caption{Effective volume and effective mass for IceCube and a detector equipped with \SI{10000}{} mDOMs. Quantities have been obtained for the detection of $e^-$ with an energy of \SI{25}{MeV}. The results for the mDOM detector include statistical uncertainties only, while the IceCube results also include systematic uncertainties. IceCube data taken from \cite{Abbasi:2011ss}.}
\label{tab:effvolume_summary}
\begin{tabular}{@{}lccc@{}}
\hline
& $\overline{V}_{\mathrm{eff}}$ per module$\mathrm{\,(\mathrm{m^3})}$ \hspace*{0.5 cm} & Total $V_{\mathrm{eff}}\,(\mathrm{m^3})$ & Total $M_{\mathrm{eff}}\,(\mathrm{Mton}$) \\ \hline
IceCube with DOMs & $\mathrm{725\pm95}$                                                 & $\mathrm{(3.8\pm0.5)\e{6}}$             & $\mathrm{3.5\pm0.5}$                      \\
Detector with \SI{10000}{} mDOMs  \hspace*{0.5 cm}& $\mathrm{1776\pm6}$                                                & $\mathrm{(1.705\pm0.005)\e{7}}$    \hspace*{0.5 cm}        & $\mathrm{15.71\pm0.05}$                       \\ \hline
\end{tabular}
\end{table*}

Table~\ref{tab:effvolume_summary} presents a comparison to the effective volumes of the IceCube detector, which was calculated in \cite{Abbasi:2011ss}. For this result, it is assumed that the \SI{10000}{} mDOMs are evenly distributed over the same depths as the IceCube modules. An increase of about a factor $2.4$ for the effective volume per module is obtained for the mDOM (in agreement with the larger effective photocathode area) resulting in a factor $\sim4.5$ larger effective volume of a detector equipped with \SI{10000}{} mDOMs compared to the current IceCube detector.

\subsection{\label{subsec:simulation}CCSN simulation}
In order to perform the simulation efficiently, positrons from IBD and electrons from ENES are generated in the ice at random positions inside the simulated volume following the model described in section \ref{subsec:SNsimulation}. It is assumed that the neutrino flux comes from the zenith. To simulate the events, a time $t$ of the flux is sampled according to the flux distribution from Fig.~\ref{fig:fluxes}. Using the associated mean energy and mean squared energy, the energy spectrum at each point in time is calculated using Eq.\,\ref{eq:nuspectra}. Afterwards, a single energy value from the spectrum is sampled according to its probability. Next, the corresponding angular cross section for this energy and the interaction that is being simulated is calculated, and a direction for the generated particle is sampled. Detected events are weighted according to their interaction probability and the total flux of (anti) neutrinos through the simulated volume. The weight $W$ is given by
\begin{equation}
    \label{eq:weights}
    W = W_{\mathrm{int}}(E) \cdot W_{\mathrm{flux}}(d) \cdot W_{\mathrm{eff}}.
\end{equation}
The first element $W_{\mathrm{int}}(E)$ represents the interaction probability of the neutrino traveling through the simulated volume. The generation volume surrounding the module is simulated as a cylinder of South Pole ice with a radius of \SI{20}{m} and a length of \SI{40}{m} with the mDOM in its center. The cylinder base is facing the SN, which simplifies the weight calculations. Thus $W_{\mathrm{int}}$ is given by
\begin{equation}
    \label{eq:weight_int}
    W_{\mathrm{int}}(E) = \sigma(E) \cdot n_{\mathrm{target}} \cdot l,
\end{equation}
where $\sigma(E)$ is the total cross section of the simulated interaction, $n_{\mathrm{target}}$ is the number of targets per volume in the ice for such an interaction, and $l=\SI{40}{m}$ is the length of the simulated volume. Note that $W_{\mathrm{int}}(E)$ is different for each simulated particle since it depends on the energy.

The flux weight $W_{\mathrm{flux}}$ is given by
\begin{equation}
    \label{eq:weight_flux}
    W_{\mathrm{flux}} =  \frac{4 \pi r^2}{N_{\mathrm{gen}}}\cdot \int \phi(t) dt= \frac{1}{N_{\mathrm{gen}}} \cdot \frac{r^2}{d^2} \cdot \int \frac{L(t)}{\mean{E(t)}} dt ,
\end{equation}
where $N_{\mathrm{gen}}$ is the number of particles that have been generated, $r=\SI{20}{m}$ is the radius of the cylindrical cap of ice facing the SN, and $d$ is the distance at which the SN is being simulated. $L(t)$ and $\mean{E(t)}$ are the fluxes and mean energies from the CCSN models.

Finally, $W_{\mathrm{eff}}$ accounts for the fact that the simulation is performed for a single mDOM at a particular depth $z_\mathrm{sim}$. Since the ice properties vary with depth, the results are scaled using the mean effective volume of the whole detector:
\begin{equation}
    \label{eq:weight_eff}
    W_{\mathrm{eff}} = N_{\mathrm{mDOM}}\cdot \frac{\mean{V_{\mathrm{eff}}(m)}}{V_{\mathrm{eff}}(m,z_\mathrm{sim})},
\end{equation}
where $N_{\mathrm{mDOM}}$ is the number of mDOMs that are being simulated and $V_{\mathrm{eff}}(m,z_\mathrm{sim})$ is the effective volume for multiplicity $m$ at the depth where the simulation is performed. We define multiplicity as the number of PMTs per module that have detected photons from the same event. The effective volume does not scale in the same way for events detected in a single PMT as it does for events detected in multiple PMTs. This is because coincident events are generally closer to the module than events detected in a single PMT, and thus are less affected by a variation in absorption length. Therefore, events are assigned a different effective volume weight $V_{\mathrm{eff}}(m)$ depending on the multiplicity $m$, with each one calculated using the method of Section~\ref{subsec:EffVol} with the appropriate multiplicity condition.

\begin{figure}[t]
    \centering
    \includegraphics[width=\columnwidth]{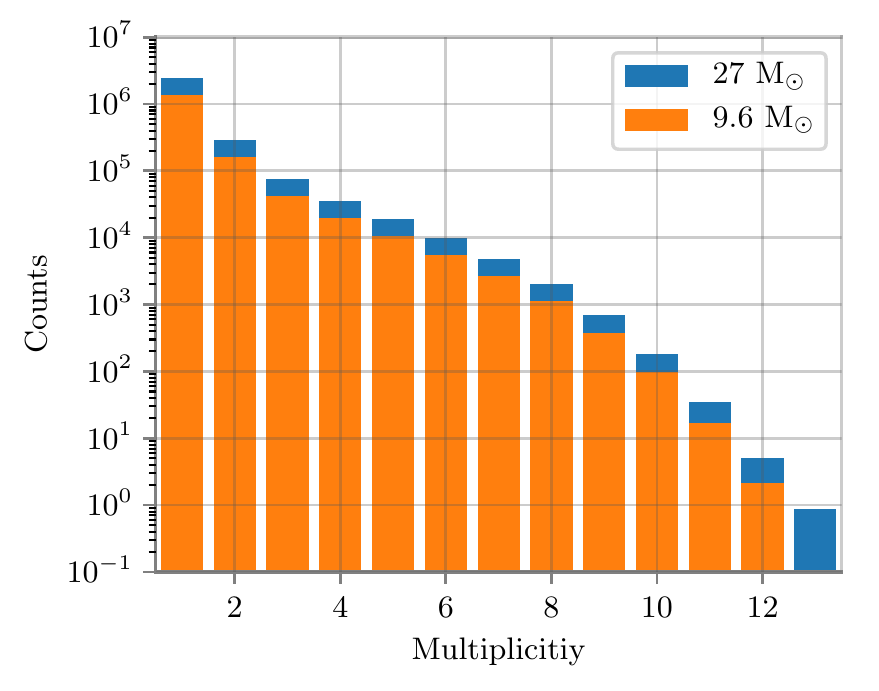}
    \caption{Multiplicity of detected events for a detector with \SI{10000}{} mDOMs for CCSNe with $27\,\mathrm{M}_{\odot}$ and $9.6\,\mathrm{M}_{\odot}$ progenitor mass at $10\,\mathrm{kpc}$.}
    \label{fig:galactic}
\end{figure}

\begin{figure}[t]
    \centering
    \includegraphics[width=\columnwidth]{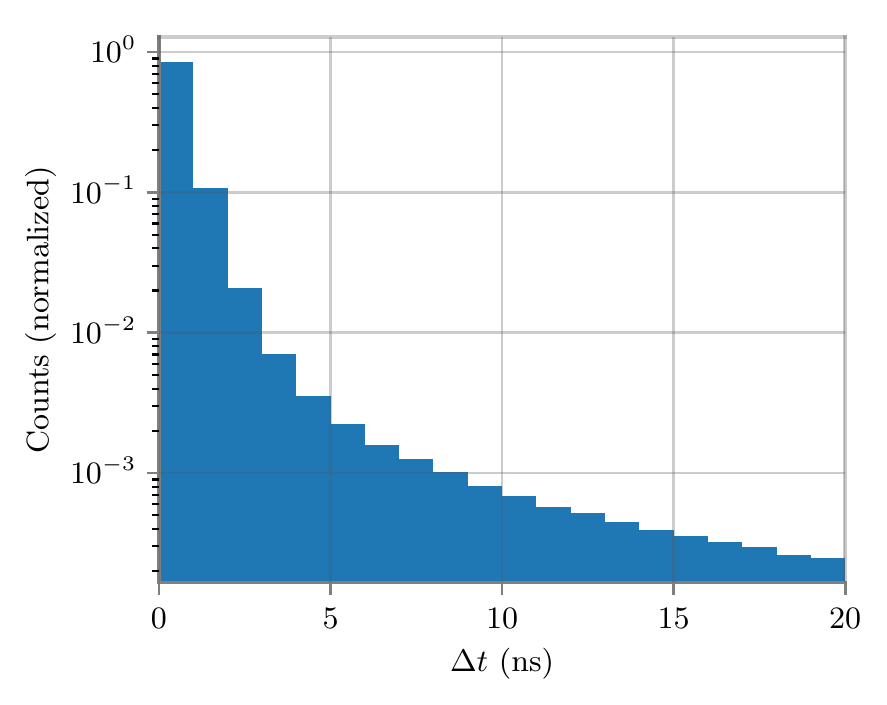}
    \caption{Time difference between first and last registered photon for events detected in more than one PMT within a single module.}
    \label{fig:timediff}
\end{figure}

The limited size of the simulated volume, chosen to speed up the simulations, does not cover the whole sensitive volume for a detection threshold of a single photon registered in a PMT. Therefore, the following calculations slightly underestimate the sensitivities. However, this does not impact the conclusions in this work, since our main results are based on events detected in coincidence in several PMTs. The distribution of event multiplicities for a CCSN simulation at \SI{10}{kpc} distance are shown in Fig.~\ref{fig:galactic}. For the high-mass CCSN we obtain more than $2.4\times10^{6}$ neutrinos with at least a single detected photon and $\sim1.4\times10^{6}$ for the low-mass progenitor. In both cases, more than $12\%$ of the detected events are in local coincidence, i.e.\ at least two photons are registered in different PMTs, while the value is slightly lower for the low-mass progenitor. Figure~\ref{fig:timediff} shows the distribution of time differences between the first and last registered photon for events detected in coincidence. It demonstrates that $\sim85\%$ of the local coincidences occur in less than $1\,\mathrm{ns}$ between the first and last registered photon, and $\sim99\%$ within $10\,\mathrm{ns}$ (note that the simulation does not include PMT pulse generation and readout). Hence, photons from the same event arrive almost simultaneously at different PMTs. This property can be exploited to suppress background and improve the identification of individual MeV neutrinos.

\section{\label{sec:identication}Identification of MeV neutrinos from CCSNe over background}
In this section, a study of local and global coincidence conditions is presented with the aim to optimize the signal-to-background ratio. 
To this end, we define trigger conditions for identifying single neutrino interactions based on fast local coincidences, as well as a minimum number of such interactions for identifying a supernova outburst. The trigger conditions are defined as follows:
\begin{itemize}
    \item[a)] Once a PMT detects a photon, a time window of length $\Delta t_{\mathrm{coin}}$ is opened. If within $\Delta t_{\mathrm{coin}}$ at least $m-1$ different PMTs in the \emph{same} module detect one or more additional photons the trigger condition is fulfilled and a \emph{neutrino event} is detected;
    \item[b)]After condition a) is satisfied, a second time window opens with $\Delta T_{\mathrm{SN}} = \SI{10}{s}$. If $N_{\nu}-1$ further neutrino events according to a) from \emph{any} mDOM in the detector fall into this time window, a supernova detection is claimed.
\end{itemize}
The trigger conditions are designed to be strict enough to effectively suppress backgrounds and keep the false detection rate at an acceptable level. In the following sections, the main background sources are discussed and an estimate of the sensitivity of the method to extra-galactic CCSNe is presented.

\subsection{\label{subsec:background}Background sources}
Due to the low radioactivity of the deep ice at the South Pole, background events originate almost exclusively from the intrinsic noise of the optical module and from the interaction of neutrinos of similar energy in the ice from other sources. These backgrounds are described in the following.

\subsubsection{\label{subsubsec:pmtnoise}PMT dark rate}
PMTs produce a measurable signal (dark rate) even if operated in complete darkness~\cite{Flyckt:712713}. The dark rate of the PMT model in the mDOM at $\SI{-30}{\celsius}$, a typical operating temperature in the deep ice, was measured with $\sim\SI{30}{s^{-1}}$ for a threshold of \SI{0.3}{pe}~\cite{Elorrieta2019}. The exact value varies from PMT to PMT and ranges up to $\sim\SI{50}{s^{-1}}$. We assume a conservative dark rate of $f_\mathrm{d}=\SI{50}{s^{-1}}$ for the purpose of this study. Following a similar calculation performed in~\cite{Boser:2013oaa}, the probability that the PMT produces a dark rate signal within a time window $\Delta t_{\mathrm{coin}}$ is given by the complementary probability of not registering any photon:
\begin{equation}
\label{eq:probnohits}
P_\mathrm{d} = 1-P\,(0,\mu=f_\mathrm{d} \Delta t_{\mathrm{coin}}) = 1-e^{-f_\mathrm{d} \Delta t_{\mathrm{coin}}}.
\end{equation}

Once a photon is detected, at least $m-1$ other PMTs in the module are required to register dark counts in the time window $\Delta t_{\mathrm{coin}}$ to satisfy trigger condition a). The probability for this can be calculated again using the complementary probability
\begin{equation}
\label{eq:probnoise}
P_{\mathrm{dark}} = 1-B_{\mathrm{cum}}\,(m-2|N_{\mathrm{PMT}}, P_\mathrm{d}),
\end{equation}
where $N_{\mathrm{PMT}}=24$ is the number of PMTs in a mDOM and $B_{\mathrm{cum}}\,(m|n,p)=\sum_{k=0}^{m}{{n}\choose{k}} p^k(1-p)^{n-k}$ is the cumulative binomial distribution for $m$ successes out of $n$ tries when the probability for success is $p$.

The rate at which trigger condition a) is satisfied due to the PMT dark rate in a module anywhere in the detector is given by
\begin{equation}
\label{eq:fnoise}
f^{\mathrm{PMT}}_{\mathrm{bg}} = P_{\mathrm{dark}}\cdot f_\mathrm{d} \cdot N_{\mathrm{tot}},
\end{equation}
with $N_{\mathrm{tot}}$ being the total number of mDOMs in the detector.

\subsubsection{\label{subsubsec:decays}Radioactive decays in the pressure vessel glass}
Scintillation and Cherenkov radiation from radioactive decays in the glass of the pressure vessel are the main background source in IceCube DOMs~\cite{Aartsen:2016nxy} and the mDOM. The radiation levels of the borosilicate glass used in the mDOM have been measured\footnote{For further information see \url{https://www.uni-muenster.de/imperia/md/content/physik_kp/agkappes/abschlussarbeiten/masterarbeiten/1712-ma_munland.pdf}.} and are listed in Table~\ref{tab:activities} assuming secular equilibrium between isotopes within the natural decay chains. Radioactive decays in other parts of the module, e.g.\ the gel, are found to be negligible. A GEANT4 simulation is performed to estimate the noise contribution of these decays in the mDOM. Within a simulation run, the number of isotopes decaying inside a time window of $\SI{20}{min}$ is drawn randomly from a Poisson distribution. All decay chains are simulated independently and the photons reaching a PMT are saved after taking its QE into account. Subsequently, the photons are mixed within the time window. Temporal correlations between mother-daughter isotopes are preserved if the decay time is within the time window. Otherwise, the decay time is changed to a random time within the window so that the number of decays remains constant (following the secular equilibrium argument). Finally, the time-ordered hits are checked for coincidences.    
\begin{table}[t]
\centering
\caption{Radioactive isotope activities in \SI{}{Bq/kg} as used in the simulation.}
\label{tab:activities}
\begin{tabular}{p{3cm}l}
\hline
    $^{235}\mathrm{U}$ chain  & $0.59\pm0.05$\\
    $^{238}\mathrm{U}$ chain  & $4.61\pm0.07$\\
    $^{238}\mathrm{Th}$ chain & $1.28\pm0.05$\\
    $^{40}\mathrm{K}$         & $61.0\pm0.9$\\
    \hline
\end{tabular}
\end{table}

\subsubsection{\label{subsubsec:solar}Solar neutrinos}
Solar neutrinos with an energy above the Cherenkov threshold in ice originate mainly from $\beta^+$-decays of $^8\mathrm{B}$ inside the Sun. The electron neutrinos, which have energies of \SI{\sim10}{MeV}~\cite{Billard:2013qya}, leave a signature in the detector similar to that of supernova neutrinos. Their spectrum is simulated in GEANT4 with a total flux of $\Phi_{\nu_e}=\SI{1.7e6}{cm^{-2}s^{-1}}$ and $\Phi_{\nu_{\mu,\tau}}=\SI{3.3e6}{cm^{-2}s^{-1}}$ using the data from \cite{Bahcall:1996qv}. We consider only the elastic scattering process with electrons in the ice, since it dominates the interaction rate by two orders of magnitude~\cite{Haxton:1998vj}.

\subsubsection{\label{subsubsec:otherbackground}Further background sources}{
The following background sources are not included in this study: 

\begin{itemize}
    \item Cosmic-ray induced atmospheric muons: these high-energy muons can penetrate into the instrumented volume, generate photons, and produce a detection pattern in a mDOM similar to that of \SI{}{MeV} neutrinos.  
    It is expected that due to their extended signature in the detector, the majority of these muons can be identified and rejected, resulting in a short dead time (not accounted for in this work). However, some of them might escape the identification algorithm and thus become part of the background for this analysis. In fact, for KM3NeT, atmospheric muons cause the majority of background events at multiplicities above six ~\cite{Aiello:2021cot}. On the other hand, the scattering length in ice is much shorter than in water. Hence, photons produced simultaneously by a muon arrive spread out in time at a module. Therefore, for our case, we expect a smaller contribution of these muons to the background. Corresponding studies require a full detector simulation, though, which goes beyond the scope of this work.
    \item Muons or electromagnetic showers from low-energy atmospheric neutrinos: these neutrinos have energies in the range of \SI{10}{MeV} -- \SI{1}{GeV}, but their interaction rate is only $\SI{\sim 1}{s^{-1}}$ in a Gton volume of ice~\cite{Salathe_2012}.
\end{itemize}
}
\begin{figure}[t]
    \centering
    \includegraphics[width=\columnwidth]{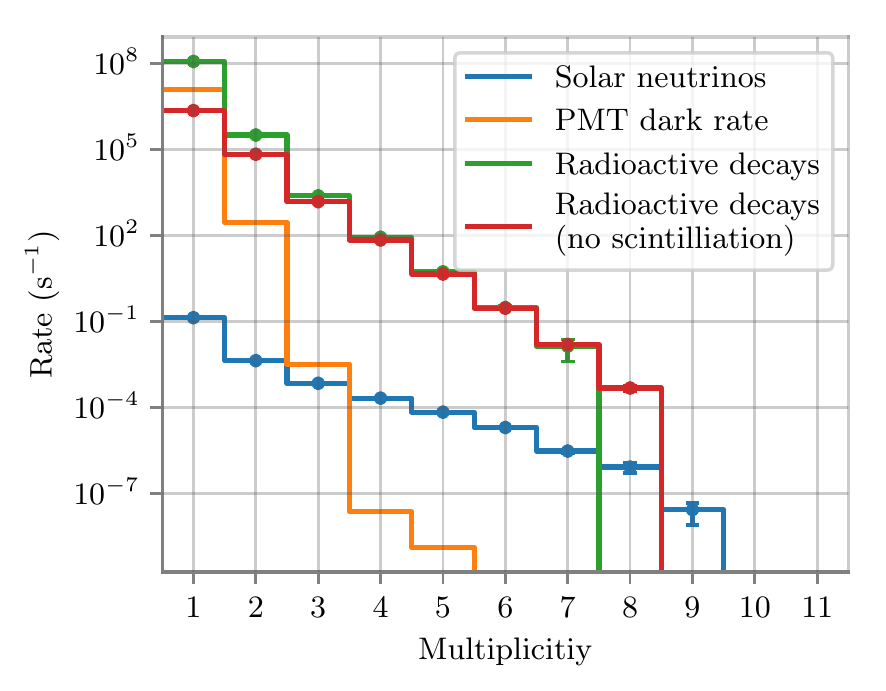}
    \caption{Multiplicity distribution produced by background in a detector of \SI{10000}{} mDOMs within a time window of \SI{20}{ns}.}
    \label{fig:noisehistogram}
\end{figure}

\begin{table*}[!ht]
\caption{Number of signal and background events and effective volume for different multiplicities (see trigger conditions in Section \ref{sec:identication}; $\Delta t_{\mathrm{coin}}=\SI{20}{ns}$, $\Delta T_{\mathrm{SN}}=\SI{10}{s}$), for a detector with \SI{10000}{} mDOMs.}
\centering
\begin{tabular}{@{}ccccc@{}}
\hline
\multirow{2}{*}{Multiplicity} & \multicolumn{2}{c}{\begin{tabular}[c]{@{}c@{}}Neutrinos from CCSN burst at $\SI{10}{kpc}$\\(counts in $\SI{10}{s}$)\end{tabular}} & \multirow{2}{*}{\begin{tabular}[c]{@{}c@{}}Background events\\(counts in $\SI{10}{s}$)\end{tabular}} & \multirow{2}{*}{\begin{tabular}[c]{@{}c@{}}Effective volume for $e^-$\\ at $\SI{25}{MeV}$ (\SI{}{m^3})\end{tabular}} \\
                                     & $27\,\mathrm{M}_{\odot}$                & $9.6\,\mathrm{M}_{\odot}$              &                                                    &                                                                                                                                  \\ \hline
$\geq5$                              & $(1.89\pm0.08)\times10^4$               & $(1.05\pm0.05)\times10^4$              & $(5.5\pm0.2)\times10^1$                            & $(5.6\pm0.2)\times10^4$                                                                                                                      \\
$\geq6$                              & $(1.00\pm0.02)\times10^4$               & $(5.5\pm0.1)\times10^3$                & $2.94\pm0.03$                                      & $(3.2\pm0.2)\times10^4$                                                                                                                      \\
$\geq7$                              & $(4.8\pm0.1)\times10^3$                 & $(2.67\pm0.07)\times10^3$              & $(1.60\pm0.06)\times10^{-1}$                       & $(2.0\pm0.1)\times10^4$                                                                                                                      \\
$\geq8$                              & $(2.03\pm0.05)\times10^3$               & $(1.12\pm0.03)\times10^3$              & $(4.9\pm0.01)\times10^{-3}$                        & $(1.1\pm0.1)\times10^4$                                                                                                                      \\ \hline
\end{tabular}
\label{tab:results}
\end{table*}

The expected background event rates for different multiplicities in a time window of $\Delta t_{\mathrm{coin}}=\SI{20}{ns}$ in a detector with \SI{10000}{} mDOMs are depicted in Fig.~\ref{fig:noisehistogram}. The largest contribution to the background comes from radioactive decays in the pressure vessel glass. Scintillation is the main background source for the overall detection rate, producing $\sim98\%$ of the detected light after a radioactive decay. Nevertheless, since this light is emitted over a long period of time, the probability for high multiplicity coincidences is low compared to that from Cherenkov emission. For $m \geq 6$, this allows the use of simulations without scintillation, which significantly reduces computational power and, as a consequence, increases statistics by a factor of ten.

A total of \SI{4500}{days} of radioactive background in a single mDOM is simulated for the no-scintillation case. Note that the cutoff at high multiplicities for radioactive and solar background in Fig.~\ref{fig:noisehistogram} is due to limitation in statistics, since these events are very rare. Due to the lack of noise events at higher multiplicities, only trigger conditions up to a threshold of $m = 8$ are considered.

In particular, the $^{232}\mathrm{Th}$ chain produces, by far, the most important contribution to the background due to $^{208}\mathrm{Tl}$, which decays into an excited state of $^{208}\mathrm{Pb}$ that produces a $\gamma$ of $\sim\SI{2.6}{MeV}$. This $\gamma$ can either interact with the vessel glass or the surrounding ice. In both cases, some of its energy is transferred to an electron, which then generates Cherenkov radiation that produces a detection pattern in the module similar to MeV neutrinos. 

Note that $\Delta t_\mathrm{coin}=\SI{20}{ns}$ is still conservative with respect to the time resolution of the module, which we expect to be around \SI{10}{ns}. As the coincidence time of most photons from supernova neutrino interactions is lower than $1\,\mathrm{ns}$ (Section~\ref{subsubsec:otherbackground}), reducing the time window to a few nanoseconds would further suppress the background. Table~\ref{tab:results} shows the number of detected events for signal and background for $\Delta t_\mathrm{coin}=\SI{20}{ns}$ and $\Delta T_{\mathrm{SN}}=\SI{10}{s}$ as well as effective detector volumes for a detector with \SI{10000}{} mDOMs. Theses conditions will also be used for the sensitivity estimation in the next chapter.

\subsection{\label{subsec:sensitivity}CCSN sensitivity}

The rate of false CCSN detections depends on the exact values of $\Delta t_{\mathrm{coin}}$, $m$, $N_{\nu}$ and $\Delta T_{\mathrm{SN}}$. If $f^{\mathrm{Solar}}_{\mathrm{bg}}$, $f^{\mathrm{PMT}}_{\mathrm{bg}}$ and $f^{\mathrm{Decays}}_{\mathrm{bg}}$ are the rates at which  trigger condition a) with parameters ($\Delta t_{\mathrm{coin}}$\,,\,$m$) is met anywhere in the detector, the total false event rate is given by
\begin{equation}
f^{\mathrm{tot}}_{\mathrm{bg}} = f^{\mathrm{PMT}}_{\mathrm{bg}} + f^{\mathrm{Decays}}_{\mathrm{bg}} + f^{\mathrm{Solar}}_{\mathrm{bg}}.
\end{equation}
Here we neglect triggers from the combination of background photons from different sources as their contribution is small. With this background, the expected number of false supernova detections $N_{\mathrm{fSN}}$ within an observation time $\delta t$ is given by
\begin{equation}
\label{eq:falseSN}
N_{\mathrm{fSN}}=f^{\mathrm{tot}}_{\mathrm{bg}} \left[1-P_{\mathrm{cdf}}\,(N_{\nu}-2,\mu=f^{\mathrm{tot}}_{\mathrm{bg}} \Delta T_{\mathrm{SN}}) \right] \delta t,
\end{equation}
where $P_{\mathrm{cdf}}$ is the cumulative Poisson distribution. 

On the other hand, the probability for a supernova at distance $d$ to produce $N_\nu$ neutrinos that meet trigger conditions a) is (note that these neutrinos are registered within $\Delta T_\mathrm{SN}$ by construction)
\begin{equation}
\label{eq:probSN}
P_{\mathrm{SN}} = 1 - P_{\mathrm{cdf}}\left(N_{\nu}-1, \mu(d)=\mu_0 (\SI{10}{kpc} / d)^2\right),
\end{equation}
where $\mu_0$ is the expected number of triggered neutrino events from the simulation of a CCSN at $10\,\mathrm{kpc}$ for the chosen multiplicity $m$ condition. 

\begin{figure}[t]
    \centering
    \includegraphics[width=\columnwidth]{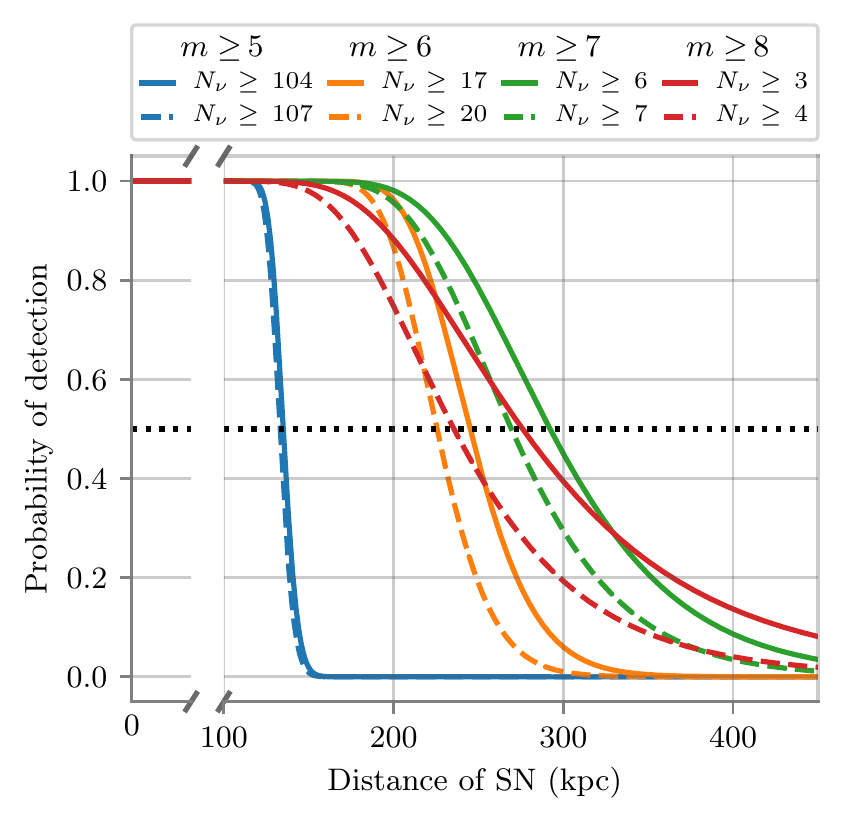}
    \caption{Probability for the detection of a CCSN of \SI{27}{M_\odot} progenitor mass as a function of distance using the trigger conditions presented in Table \ref{tab:falseSN}.}    \label{fig:prob_detection_distance}
\end{figure}

Table~\ref{tab:falseSN} presents trigger conditions for which the false CCSN detection rate per year ($\delta t = \SI{1}{year}$) is $\leq\!1$ and $\leq\!0.01$, respectively. The distance at which a CCSN is still detectable with a probability of 50\% for such conditions is also listed. Due to the high rate of low multiplicities from background sources, the best results are obtained for high multiplicities, with the largest distance reached at $m\geq7$. The trigger condition ($m\geq7$\,,\,$N_{\nu}\geq7$) can be used to send supernova alerts with very high confidence (about one false detection per century), and identify CCSN at a distance of \SI{269}{kpc} with $50\%$ probability. With a relaxed set of conditions of ($m\geq7$\,,\,$N_{\nu}\geq6$), SNe up to $291\,\mathrm{kpc}$ can be detected with less than one false CCSN detection per year. A detailed discussion of acceptable false detection rates goes beyond the scope of this paper but the assumed rates are rather conservative according to \cite{Kharusi:2020ovw}. Figure~\ref{fig:prob_detection_distance} shows the probability from Eq.~\ref{eq:probSN} as a function of distance that a \SI{27}{M_\odot} CCSNe is detected for the aforementioned cases. These results clearly demonstrate the potential of the method to increase the sensitivity to distant CCSNe compared to the current IceCube detector, which can detect CCSNe up to $\SI{50}{kpc}$ with a false detection rate of $\SI{0.1}{{year}^{-1}}$~\cite{Abbasi:2011ss}.

\begin{table}[t]
\caption{False CCSN detection rate and range of supernova detection (50\% probability) for different values of $m$ and $N_{\nu}$ (see trigger conditions in Section \ref{sec:identication}; $\Delta t_{\mathrm{coin}}=\SI{20}{ns}$, $\Delta T_{\mathrm{SN}}=\SI{10}{s}$).}
\begin{tabular}{lccc}
\hline
\multicolumn{2}{c}{Trigger}          & \multicolumn{1}{c}{False CCSN rate} & \multicolumn{1}{c}{Range (\SI{}{kpc})}             \\
$m$      & $N_{\nu}$ & \SI{}{(year^{-1})}             & $27\,\mathrm{M}_{\odot}$ ($9.6\,\mathrm{M}_{\odot}$) \\ \hline
\multirow{2}{*}{$\geq5$} & $\geq104$  & $0.7$                          & $135$ ($101$)                                 \\
                         & $\geq107$  & $<0.01$                        & $133$ ($99$)                                 \\ \hline
\multirow{2}{*}{$\geq6$} & $\geq17$  & $0.9$                          & $245$ ($182$)                                 \\
                         & $\geq20$  & $<0.01$                        & $225$ ($167$)                                 \\ \hline
\multirow{2}{*}{$\geq7$} & $\geq6$   & $0.4$                          & $291$ ($216$)                                 \\
                         & $\geq7$   & $0.01$                         & $269$ ($200$)                                 \\ \hline
\multirow{2}{*}{$\geq8$} & $\geq3$   & $0.2$                          & $275$ ($204$)                                 \\
                         & $\geq4$   & $<0.01$                        & $235$ ($174$)                                 \\ \hline
\end{tabular}
\label{tab:falseSN}
\end{table}

In the case that other telescopes detect a CCSN, we can also analyse archival data in search for a signal. Let us assume that the time at which the neutrinos should have arrived at the detector is known with an accuracy of one hour~\cite{Cowen:2009ev}. Then Eq.~\ref{eq:falseSN} gives the expected number of false supernova detections $N_{\mathrm{fSN}}$ generated by the background within $\delta t = \SI{1}{h}$. The probability $Q$ that the background did not produce a false supernova-like signal within this interval is \begin{equation}
    \label{eq:probFalseSignal}
    Q = P(0, N_{\mathrm{fSN}}(\delta t = \SI{1}{h})).
\end{equation}

\begin{figure}[t]
    \centering
    \includegraphics[width=\columnwidth]{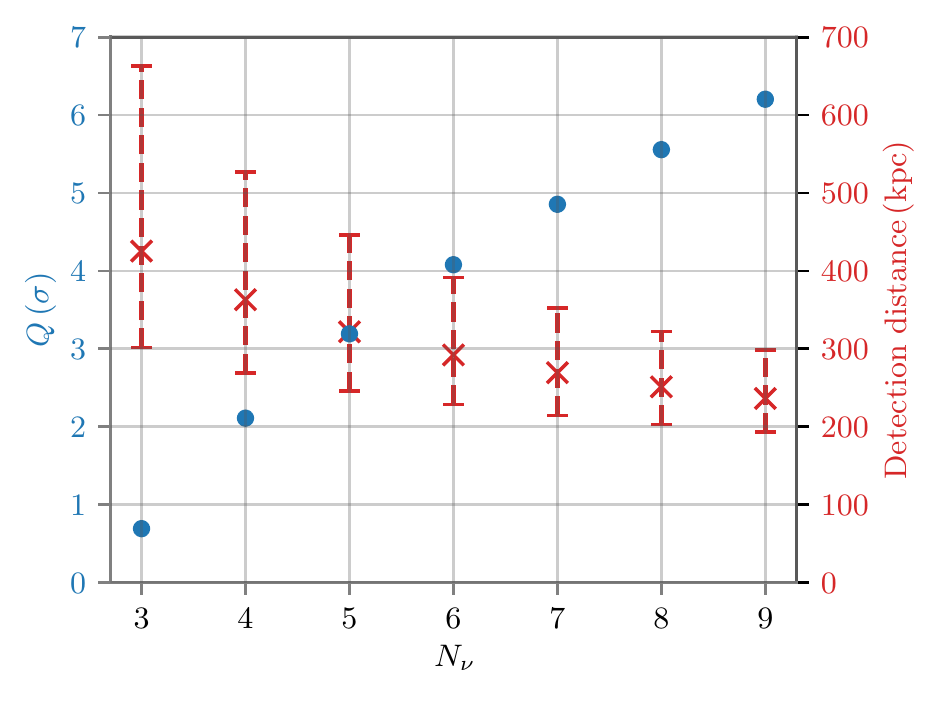}
    \caption{Detection prospects for a CCSN whose time is known to within \SI{1}{h} ($m\geq7$\,,\,$\Delta t_{\mathrm{coin}}=\SI{20}{ns}$\,,\,$\Delta T_\mathrm{SN} = \SI{10}{s}$). \textbf{Left axis:} probability in $\sigma$ that the signal is not produced by background fluctuations. \textbf{Right axis:} distance at which a $27\,\mathrm{M}_{\odot}$ progenitor mass CCSN is detected with $10\%$ (upper boundary), $50\%$ (middle mark) and $90\%$ probability (lower boundary), when at least $N_{\nu}$ detected events are required.}
    \label{fig:significance}
\end{figure}

Figure~\ref{fig:significance} shows $Q$ in terms of one-sided Gaussian standard deviations for the detection of a $27\,\mathrm{M}_{\odot}$ CCSN as a function of the minimum number of neutrino events $N_\nu$ together with the distances at which such a CCSN can be detected. Only the case $m\geq7$ is shown since it provides better results than other multiplicity conditions. $Q$ drastically increases with the increase of $N_\nu$ while the range for identifying distant supernovae decreases only moderately. For example, for a number of detected events $N_\nu=5$ a background origin can be excluded at $\SI{3.2}{\sigma}$, while at least a corresponding number of events will be detected in $50\%$ of cases from a $27\,\mathrm{M}_{\odot}$ CCSNe at a distance of $\SI{322}{kpc}$. If $N_\nu=7$ events with $m\geq7$ are detected we obtain a $\SI{4.9}{\sigma}$ confidence that such signal was not produced by background with a $50\%$ detection probability at $\SI{269}{kpc}$ distance.

The KM3NeT collaboration estimates a $5\,\sigma$ discovery range of 30 kpc for a CCSN with a $27\,\mathrm{M}_{\odot}$ progenitor star \cite{Aiello:2021cot} if the arrival time of the burst is exactly known\footnote{Notice that we use a different model despite the progenitor mass being the same; KM3NeT uses fluxes from 3D explosions models \cite{Tamborra:2013laa}, limiting the simulation time to $\sim\SI{500}{ms}$, while we use 1D models \cite{Sukhbold:2015wba} which should be precise enough for the aim of this study.}. We can make the same assumption and, following their approach, obtain the sensitivity for such a detection from ${Z=\sqrt{2((s+b)\mathrm{ln}(1+s/b-s)}}$. Here, $s$ is the expected number of signal and $b$ the expected number of background events. The $5\sigma$ discovery horizon in this scenario reaches $\SI{315}{kpc}$ for a $27\,\mathrm{M}_{\odot}$ CCSN using $m\geq7$, and $\SI{234}{kpc}$ for the $9.6\,\mathrm{M}_{\odot}$ model. It should be noted that the KM3NeT result is based on a full detector simulation with atmospheric muon background rejection while this study does not. As stated before, we expect the contribution of the muon background in ice to be less significant than in water at high coincidence multiplicities. The trigger conditions applied in this work can also be further optimized, e.g., by reducing $\Delta t_{\mathrm{coin}}$ or by shortening the $\Delta T_{\mathrm{SN}}$ window. The latter is based on the fact that the flux is expected to be most intense in the first few seconds of the burst. 

In this work, we have used the detection horizon as a performance metric. The ranges achieved reach up to several hundred kpc. However, most CCSNe candidates within this range are located within $\sim\SI{50}{kpc}$ in the Milky Way or the Small and Large Magellanic Clouds. Beyond, CCSN progenitors are expected again in the nearest galaxy, M31, which is  $\sim\SI{800}{kpc}$ away, and in more distant galaxies. Using the estimated CCSNe population based on recent observations and scaled to the star formation rate \cite{Boser:2013oaa}, we obtain a CCSNe detection rate of $\SI{0.046}{year^{-1}}$, i.e.\ one SN about every \SI{20}{years} ($m\geq7$ \,,\,$N_{\nu}\geq7$). One way to increase the detection horizon and thus the SN detection rate is to use a pressure vessel with lower radioactive contamination. In fact, this is not an unrealistic scenario, as pressure vessels for IceCube-Gen2 are currently under investigation with radioactive contamination levels about a factor of 100 lower than those assumed in this work.

\begin{figure}[t]
    \centering
    \includegraphics[width=\columnwidth]{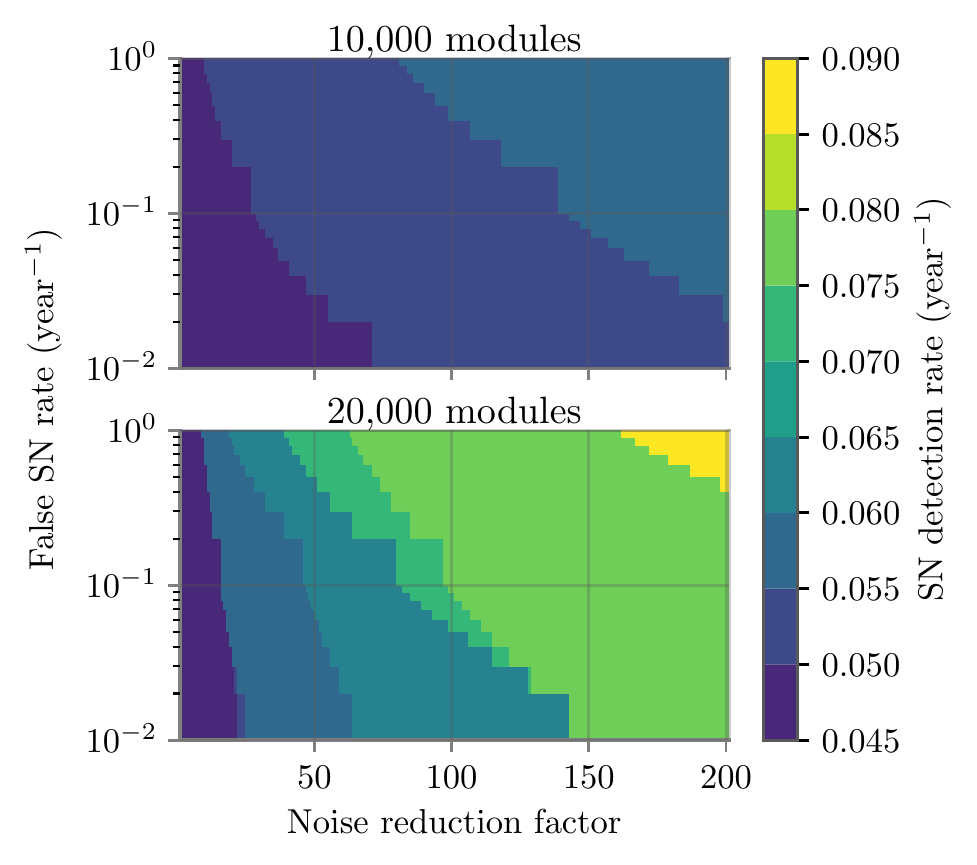}
    \caption{CCSN detection rate for hypothetical detectors with \SI{10000}{} (\textbf{upper}) and \SI{20000}{} (\textbf{lower}) mDOMs as a function of the false SN detection rate and a reduction in radioactive noise compared to standard mDOMs. The CCSN detection rates have been calculated using the estimated CCSNe population from \cite{Boser:2013oaa} based on actual observations and scaled to the star formation rate.
    }
    \label{fig:detectionrate}
\end{figure}

Figure~\ref{fig:detectionrate} shows the expected SN detection rate for hypothetical detectors with \SI{10000}{} and \SI{20000}{} mDOMs as a function of the false SN rate and the noise reduction factor. In order to double the CCSN detection rate, the noise level must be significantly reduced and the number of modules increased. 
Under the challenging scenario that the radioactive noise of the vessel glass could be reduced by a factor of $\sim 140$, a doubling of the number of installed modules would allow the false SN detection rate to be kept below $\SI{0.01}{year^{-1}}$ while doubling the expected CCSN detection to one every ${\sim}12$~years.
The same would be achieved with a reduction factor of ${\sim}100$ and a relaxation of the trigger conditions allowing a false detection rate of $\SI{0.1}{year^{-1}}$.

\section{\label{sec:conclusions}Conclusions}

In this paper, we have shown that exploiting temporal coincidences between detected photons within a segmented photosensor significantly increases the sensitivity of sparsely instrumented neutrino telescopes to MeV neutrinos and thus distant CCSNe. Due to its negligible optical background, the deep ice at the South Pole is particularly well suited for this purpose. For a detector equipped with \SI{10000}{} sensors consisting of 24 3-inch photomultipliers, we find that CCSNe up to a distance of \SI{269}{kpc} can be identified with $50\%$ probability with $0.01$ false SN detection per year.

Increasing the number of installed modules to \SI{20000}{} and using pressure vessels with significantly reduced optical background could extend the range such that one CCSN every ${\sim}12$~years can be observed. If the arrival time of CCSN neutrinos is known from an independent observation with $\delta t = \SI{1}{h}$, a $27\,\mathrm{M}_{\odot}$ CCSN at [322, 269]\,kpc can be detected in $50\%$ of cases and with a [3.2, 4.9]\,$\sigma$ certainty that the signal was not produced by background. We note that our studies, which are based on the simulation of a single sensor, do not account for background from atmospheric muons. However, we expect that this background can be effectively suppressed with tailored selection and reconstruction algorithms.

Since each sensor represents a self-contained detector for MeV neutrino, the sensitivities depend only on the number of photosensors but \emph{not} on the inter-module spacing. Hence, future sparsely instrumented neutrino telescopes like IceCube-Gen2, optimized for TeV--PeV neutrinos, will not only increase the sensitivity to the high energy universe but, if equipped with segmented photosensors, also significantly increase our reach for the detection of CCSNe with MeV neutrinos beyond the galactic range.

\vspace{3mm}
{\small
\noindent \textbf{Acknowledgments:}  We kindly thank \emph{T.~Hanka} for providing the data for the CCSN neutrino fluxes and Segev BenZvi for valuable comments on the manuscript. This work was supported by the German Bundesministerium für Bildung und Forschung (BMBF) Verbundforschung grant 05A20PM2.

}

\bibliographystyle{spphys}
\bibliography{main.bib}

\end{document}